\documentclass[amssymb, amsmath, accepted=2019-05-01]{quantumarticle}

\usepackage{graphicx}
\usepackage[numbers, sort&compress]{natbib}
\usepackage{hyperref}

\begin{document}

\title{Multiplexed Quantum Random Number Generation}%

\author{Ben Haylock}%
\thanks{This author contributed equally to this work}
\orcid{0000-0002-3925-4484}
\affiliation{Centre for Quantum Dynamics, Griffith University, Brisbane, 4111, Australia}
\author{Daniel Peace}
\thanks{This author contributed equally to this work}
\affiliation{Centre for Quantum Dynamics, Griffith University, Brisbane, 4111, Australia}
\author{Francesco Lenzini}
\affiliation{Centre for Quantum Dynamics, Griffith University, Brisbane, 4111, Australia}
\author{Christian Weedbrook}
\affiliation{Xanadu, 372 Richmond St. W., Toronto, M5V 2L7, Canada}
\author{Mirko Lobino}
\email{m.lobino@griffith.edu.au}
\affiliation{Centre for Quantum Dynamics, Griffith University, Brisbane, 4111, Australia}
\orcid{0000-0001-8091-8951}
\affiliation{Queensland Micro- and Nanotechnology Centre, Griffith University, Brisbane, 4111, Australia}
\date{\today}%
\begin{abstract}
Fast secure random number generation is essential for high-speed encrypted communication, and is the backbone of information security. Generation of truly random numbers depends on the intrinsic randomness of the process used and is usually limited by electronic bandwidth and signal processing data rates. Here we use a multiplexing scheme to create a fast quantum random number generator structurally tailored to encryption for distributed computing, and high bit-rate data transfer. We use vacuum fluctuations measured by seven homodyne detectors as quantum randomness sources, multiplexed using a single integrated optical device. We obtain a real-time random number generation rate of 3.08 Gbit/s, from only 27.5 MHz of sampled detector bandwidth. Furthermore, we take advantage of the multiplexed nature of our system to demonstrate an unseeded strong extractor with a generation rate of 26 Mbit/s.
\end{abstract}
\maketitle
\section{Introduction}
Information security\cite{Ware:67} is a foundation of modern infrastructure with quantum optics set to play a prevalent role in the next generation of cryptographic hardware\cite{Zhang:14}. Randomness is a core resource for cryptography and considerable effort has gone into making systems suitable for supplying high bit rate streams of random bits. The randomness properties of the source have a profound effect on the security of the encryption, with several examples of compromised security from an attack on the random number generator \cite{Debian,Dorrendorf:09,Nohl:08}. In this area, quantum optics has provided advantages over previous methods, enabling random number generation with high speeds and enhanced security \cite{H-C:17,Ma:16,Hart:17}.

The gold standard for security in random number generators comes from device independent quantum random number generators (QRNGs) \cite{Nie:16}, where the output is certified as random regardless of the level of trust in the generator. These generators require an experimental violation of a Bell-type inequality, an extremely difficult task, limiting generation rates to well below practical requirements ($<$kbit/s) \cite{Pironio:10,Bierhorst}.  Other approaches based on the Kochen-Specker theorem to prove value indefiniteness of the measurement, have demonstrated faster but not yet usable generation rates (25kbit/s) \cite{Kulikov:17,Abbott:12}. Recent advances\cite{Liu:18} have built upon the notion of Bell inequality violations using device independent quantum random number generators. Remarkably, these allow the closure of any security loopholes related to the how the device is made, i.e., independent of the device implementations. Currently, high-speed (Mbit/s-Tbit/s) quantum random number generation relies on trusted or semi-trusted generators, where the independence of the randomness from classical noise is experimentally tested \cite{Gabriel:10,Li:14,Lunghi:15,Mitchell:15,Marangon:2017,Virte:14,Wayne:09,Xu:12}. While these systems have no quantum physical guarantee of their randomness, they are usually denoted as QRNGs due to the quantum mechanical origin of the randomness. This trade-off between speed and security is mostly caused by the experimental complexity of fully secure implementations.

Entropy sources sufficient for randomness generation rates up to 1.2Tb/s have been demonstrated\cite{Sakuraba:15}, However, these systems are not capable of real time random number generation at full speed due to processing bandwidth limitations, instead performing off-line processing on captured data to generate randomness. Thus these schemes are not suited for providing high-speed random numbers for cryptography. Regardless of generation procedure, implementations of randomness extraction are limited by electronic logic speeds or detection bandwidths. Therefore, for high speed real time random number generation, the most feasible solution is to create many parallel sources with lower entropy rates.

Multiplexed quantum random number generation has previously been theoretically proposed as a solution to post-processing bottlenecks in the real-time rate of QRNGs \cite{Hart:17,Ma:16}. Previous demonstrations of a parallel nature remain focused on increasing single-channel data rates. Gr\"{a}fe et al. extend single photon path-encoding based QRNGs to the multi-mode case increasing the bitrate for a fixed measured photon flux\cite{Grafe:14}. Haw et al. sample two separate frequency slices of their homodyne detector bandwidth in a vacuum fluctuation QRNG, enabling them to generate randomness at twice their digitization rate\cite{Haw:15}. Both demonstrations remain ultimately rate-limited by generation\cite{Grafe:14} or detection\cite{Haw:15} rates. The goal of multiplexing should be to increase the data rate regardless of single channel bandwidth limitations as shown by IDQuantique by multiplexing together four separate devices on a single interface\cite{IDQuantique}. 

A multiplexing architecture is more versatile than increasing the rate of a single data stream, allowing the use of more complex extraction techniques and randomness distribution at rates faster than a single channel capacity. Randomness extractors are algorithms which, given a bit string from a weakly random physical source, produce a shorter sequence of truly random bits\cite{Shaltiel:11}. Previous works have largely used seeded extractors, which require a uniform random seed to convert the input to random bits. Such extraction is often described as randomness expansion, as it cannot extract true randomness without already having some at the input\cite{H-C:17}. Seeded extractors rely on the uniformity of the seed and independence of the output from this seed for security. We can relax both of these requirements with a multi-source extractor. 

A single random output is produced from two weakly random inputs in a multi-source extractor. The main advantage of this approach is that random numbers can be generated without any initial random seed. Many examples of multi-source extractors exist that allow for unseeded extraction with low entropy loss, including constructions which are strong extractors in the presence of quantum side information\cite{Kasher:10}. 

We experimentally demonstrate a high-speed parallel quantum random generator whose total rate is not limited by generation or detection rates but rather by the number of parallel channels used. While Gbps real-time generation rates have been previously been demonstrated\cite{Ugajin:17,Zhang:16,Marangon:18}, our scheme provides a simple method scheme to overcome electronic and processing bandwidth limitations. 

\section{Random Number Generation Scheme}
\begin{figure}
	\includegraphics[width=0.49\textwidth]{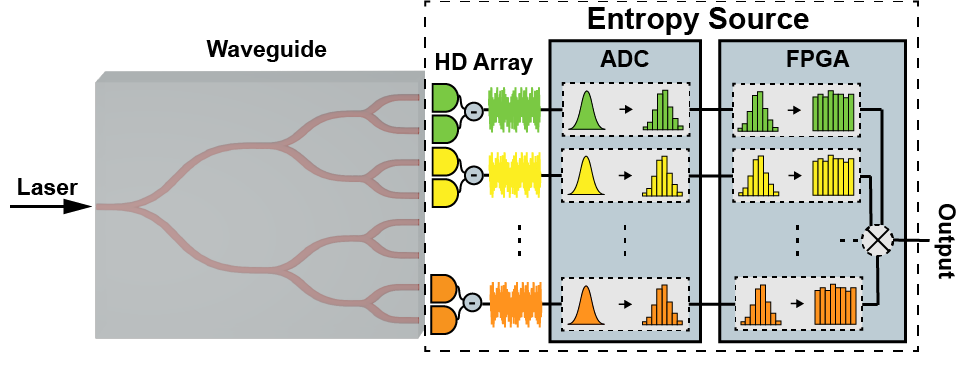}
	\caption{Scheme for multiplexed quantum random number generator based on quadrature measurements of the vacuum state. A low noise, Koheras Boostik laser at 1550nm is coupled in and out of a lithium niobate waveguide network through butt-coupled fiber arrays. Light from the outputs is sent into seven homodyne detectors. The detector signals are sent to the ADC and FPGA for digitization, processing and randomness extraction. This schematic shows four channels, the experimental implementation used up to seven channels.}
	\label{fig1}
\end{figure}

\begin{figure}
	\includegraphics[width=0.49\textwidth]{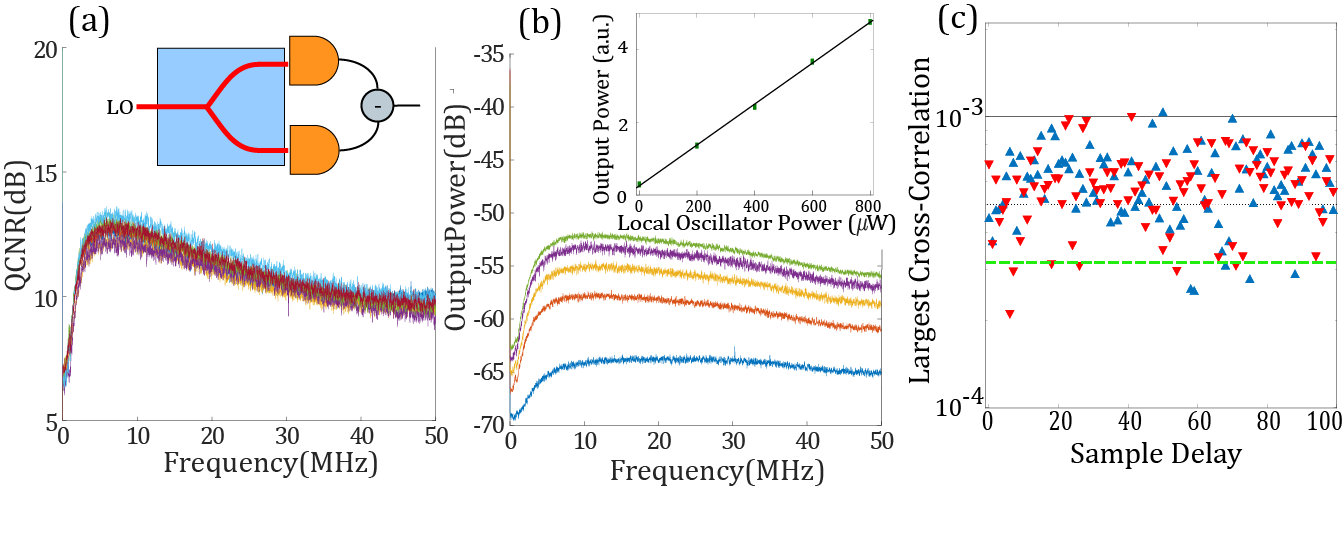}
	\caption{(a) Quantum to classical noise ratio (QCNR) of all seven homodyne detectors used, with inset showing a schematic description of a homodyne detector, orange hemispheres represent photodiodes. Power used is 1mW and all powers are measured at the output of the waveguide device. (b) Linearity of homodyne detector response with increasing local oscillator power, shown using a representative (channel 4). Blue - 0$\mu$W, orange -200$\mu$W, yellow - 400$\mu$W, purple - 600$\mu$W, green - 800$\mu$W.  Inset, a plot of the linear response at 5MHz. (c) The largest positive (blue) and negative (red) correlations between any pairwise combinations of eight-bit encoded homodyne measurements from the seven channels. The green line represents the ideal value for the size of the data set (10 million samples).}
	\label{fig2}
\end{figure}

A schematic of the multiplexed QRNG scheme is shown in Figure \ref{fig1}. The noise source of each channel of our multiplexed design comes from homodyne measurements of vacuum state \cite{Gabriel:10,Shen:10} (see inset in Figure \ref{fig2}(a)). A laser is sent onto a 50-50 beamsplitter while vacuum enters the other port, subsequently the two outputs of the beamsplitter are detected on two photodiodes and the difference between the two photocurrents is amplified. The homodyne current is proportional to a measurement of the quadrature operator of the vacuum state, and its value is independent and unpredictable within a Gaussian distribution with zero mean. 

The homodyne detectors (HDs) used in this demonstration follow the design of Kumar et al.\cite{Kumar:12} and have an electronic bandwidth of ~100MHz. The quantum efficiencies of the photodiodes used (PD20, oemarket) are $>$67\% (70\% typical). The quantum signal to classical noise ratio (QCNR) is defined as QCNR$=10\log_{10}(\sigma_Q^2/\sigma_E^2)$ where $\sigma_Q^2$ is the quantum noise variance, and $\sigma_E^2$ is the classical noise variance. Using the equation $\sigma_Q^2=\sigma_M^2-\sigma_E^2$, QCNR can be calculated from experimental measurement of the variance of the output of the homodyne detector with the local oscillator off ($\sigma_E^2$), and with the local oscillator on ($\sigma_M^2$). Figure \ref{fig2}(a) shows the results for all seven homodyne detectors. We see that for all channels of our design this ratio exceeds 10dB across 30 MHz, with a measured common mode rejection ratio of $>$27dB across all seven detectors. To confirm that our detectors were measuring vacuum fluctuations, we determine the linearity of the noise as a function of the laser power (see Figure \ref{fig2}(b)). As the QCNR is the ratio between this response and the value at zero power (blue trace), it also scales linearly with input power.  The independence of the outcomes of each of the channels was verified from cross-correlation measurements shown in Figure \ref{fig2}(c). The cross-correlation for ideal uniform data is  $1/\sqrt{n}=3.16\times 10^{-4}$. All 21 channel pairing cross-correlations lie between $2.83\times 10^{-4}$ and $3.51\times 10^{-4}$.

Integrated optics provides a compact and stable way to implement the set of beamsplitters needed to feed many homodyne detectors. We fabricate a 1:32 multiplexer using annealed proton exchanged waveguides in lithium niobate with a device footprint of 60mm x 5mm\cite{Lenzini:15}. The device has insertion losses of $\approx$7 dB ($\approx$22dB total loss per channel) and we choose balanced outputs to send to the seven homodyne detectors. Total input power to the array is approximately 160mW ($\approx$1mW per channel at output), which will ultimately limit the maximum number of channels. This limitation may be overcome by using several multiplexed lasers as the input. Our choice of lithium niobate is designed for a semi-integrated system, where a compact butterfly diode laser, the waveguide device, a butt coupled linear photodiode array, and all the electronics could be housed on a single circuit board. 

Several data processing steps are implemented in order to transform the analog signals from the homodyne detectors into a stream of random bits (see Fig. 1). First, the analog output of each detector is digitized into 12 bits per sample using an analog to digital converter (ADC, Texas Instruments ADS5295EVM). In our demonstration no anti-aliasing filter is used and as such frequencies higher than the detection bandwidth contribute to the signal. The digitized results from each outcome are sent in parallel into a field programmable gate array (FPGA, Altera Arria II GX Development Kit) for the remainder of the randomness extraction protocols. If the outputs are to be multiplexed back together rather than used in parallel, multiplexing occurs after the randomness extractor. Multiplexing is done by interleaving channel by channel and the output of the extractors is written to the on-chip memory of the FPGA and may be transferred to a computer via a USB cable for randomness testing. Randomness verification tests are performed off-line by transferring experimental data to a PC. In any application the memory connected to the FPGA is equally suited to storing and sending the randomness as the memory of a PC.

Three different extraction methods are demonstrated that convert the unpredictable measurement outcomes of the homodyne detectors into random bit streams. In the first extractor (A), which we call `raw bit extraction’, we take the eight least significant bits (LSBs) from the ADC and discard the remaining 4 bits per sample. This extractor follows the design of the 'environmental immunity' procedure of \cite{Symul:11}. This extractor is designed to minimise the influence of the classical noise on the output signal. Demonstration of the security of this protocol is through the high entropy of the experimental output, rather than from any theoretical proof.

The second extractor (B) is based on the second draft of NIST Special Publication 800-90B\cite{Sonmez:16}. The authors list a set of vetted randomness extractors, one of which is the keyed algorithm CMAC (Cipher-based Message Authentication Code)\cite{Dworkin:05} with the AES (Advanced Encryption Standard)\cite{NIST:01} block cipher. For an input with k bits of min-entropy i.e., one where the maximum probability of any outcome is bounded by $2^k$ , when $\leq k/2$ bits are taken from the 128 bit output of the extractor, full-entropy output bits are produced \cite{Sonmez:16}. The remaining $\lfloor128-k/2\rfloor$ bits are used to refresh the seed. We take eight LSBs from sixteen consecutive digitization samples to form the 128-bit input to each run of the extractor. The AES hash is implemented on the FPGA using the TinyAES core\cite{Hsing}.

The third extractor(C) takes advantage of the fact that we have many independent sources, and as such can use a multi-source extractor. Examples of both weak (seed-dependent, non-reusable seed) and strong (seed-independent, reusable seed) extractors have been shown for QRNGs. The security of these extractors relies on the quality of the previously created random seed. Multi-source extractors discard the necessity for a truly random seed. Instead, they take two or more partially random bit-strings from weak randomness sources and produce a truly random output. Given sufficient randomness of the inputs, a strong multi source extractor outputs bits that are uncorrelated with any of the inputs, providing randomness even with full knowledge of all but one of the inputs. We implement a single bit two-source extractor, as described in \cite{Bouda:12}, which produces a uniform output providing at least one of the inputs has more min-entropy than the output number of bits. Each extractor takes two 36-bit strings from two different homodyne detectors, each consisting of three 12-bit samples. As such we need an even number of input channels for this extractor, and using six of the detectors we create three of these extractors and multiplex the outputs together.

\section{Entropy Source Evaluation}
We first evaluate the worst-case conditional min-entropy of the 12-bit output of the ADC for each channel to find the amount of entropy sourced from the measurement of the vacuum state. Using the procedure described by Haw et al. \cite{Haw:15} for the worst-case conditional min-entropy ($H_{min}$) we can find a lower bound for the maximum extractable randomness given the discretized measured distribution $M_{dis}$, conditioned on the classical side information $E$ \cite{Haw:15,Renner:08}: 
\begin{equation}
H_{min}(M_{dis}|E)=-\log_2 \left[\max_{e\in \mathbb{R}} \max_{m_i\in M_{dis}} P_{M_{dis}|E}(m_i|e)\right]
\label{minent}
\end{equation}
where $m_i$ and $e$ are the samples from their respective distributions. Taking a maximum classical noise spread of $e_{max}=5\sigma_E$, and using a representative sample of $1\times 10^6$ samples per channel we numerically evaluate Equation \ref{minent} and find a worse case min entropy of $H_{min}(M_{dis}|E)\geq 9.201$ across all seven channels, for a near optimal digitization range. This value describes the entropy component immune to classical noise sources; however, it does not describe the component immune to quantum side information, as the extractors we use are not secure against such attacks.

\begin{table}
	\caption{Summary of results for our three constructions. The bitrate can be found by multiplication of sampling rate, number of extractors, and extracted bits per sample.The min-entropy per 8 bits for extractor C can be obtained from the reported min-entropy per bit = 0.986 multiplied by eight.}
	\begin{tabular}{p{17mm}|c|c|c}
		& Raw 8 Bit & AES &	Two Source\\ \hline
		Extractor Channels & 7 & 7 & 3\\ \hline
		Sampling Rate (MSPS) & 55 & 50 & 52\\ \hline
		Bits per Sample & 8 & 8 $\times$ 63/128 & 12 $\times$1/72\\ \hline
		Generation Rate (Gbps) & 3.08 &	1.37 & 0.026 \\\hline
		Min-Entropy per 8 bits &	7.897 &	7.902 &	7.890\\ \hline
		IID Test\cite{Sonmez:16} &	Pass &	Pass &	Pass\\\hline
		Randomness Test\cite{Rukhin} &	Pass &	Pass &	Pass\\
	\end{tabular}
	\label{table}
\end{table}

If each sample in a test set from a noise source is mutually independent and have the same probability distribution, that noise source is considered to be independent and identically distributed (IID). The NIST SP800-90B entropy assessment package \cite{github} uses a range of statistical tests to attempt to prove that a sample is not IID. If none of the tests fail, the noise source is assumed IID. The output entropy of the raw bit extraction (A) is tested using the entropy estimate procedure from NIST SP800-90B, and find the sample passes the IID test with an entropy of 7.897 bits. This extractor is designed to minimise the effect of the classical noise on the output signal. The total bit rate of this construction is given by the product of the sample rate (55MSPS), extracted bits per sample (8), and number of channels (7), and is 3.08 Gbit/s. We sample at least 27.5 MHz of the homodyne detector bandwidth as allowed by the Shannon-Hartley limit\cite{Hart:17}. The sampling rate is limited by the interface between the ADC and FPGA. Thus, we generate 112 Mbit/s per MHz of single channel detector bandwidth. We note that previous implementations have sampled more than an order-of-magnitude more detector bandwidth with superior detectors and digitization \cite{Haw:15}, which will enable parallel QRNG from vacuum to reach much faster rates than in this demonstration. 

Using this entropy estimate of the 8-bit raw data we construct a vetted CMAC keyed extractor (B), taking 63 out of 128 bits of the output to ensure the number of bits we use is less than half the input entropy. Entropy tests of the output give a min-entropy of 7.902 bits and the sample passes the IID test. Finally, we measure the output entropy of our two-source extractor(C) to be 0.986 bits as it is a single bit extractor, and it also passes the IID test. Whilst the output bit rate is much lower because it requires two 36 bit inputs (72 raw bits total) to produce one output bit, it removes the necessity for an externally generated seed, prevalent in most QRNG demonstrations. The results for all three extractors are summarized in Table \ref{table}. The NIST statistical test suite [37] is also used to identify any statistical correlations that may make the data non-random. We run the test suite on each of our constructions over a minimum sample size of  $7\times 10^8$ bits and find extraction methods A, B, and C pass all tests.

\begin{figure}
	\includegraphics[width=0.49\textwidth]{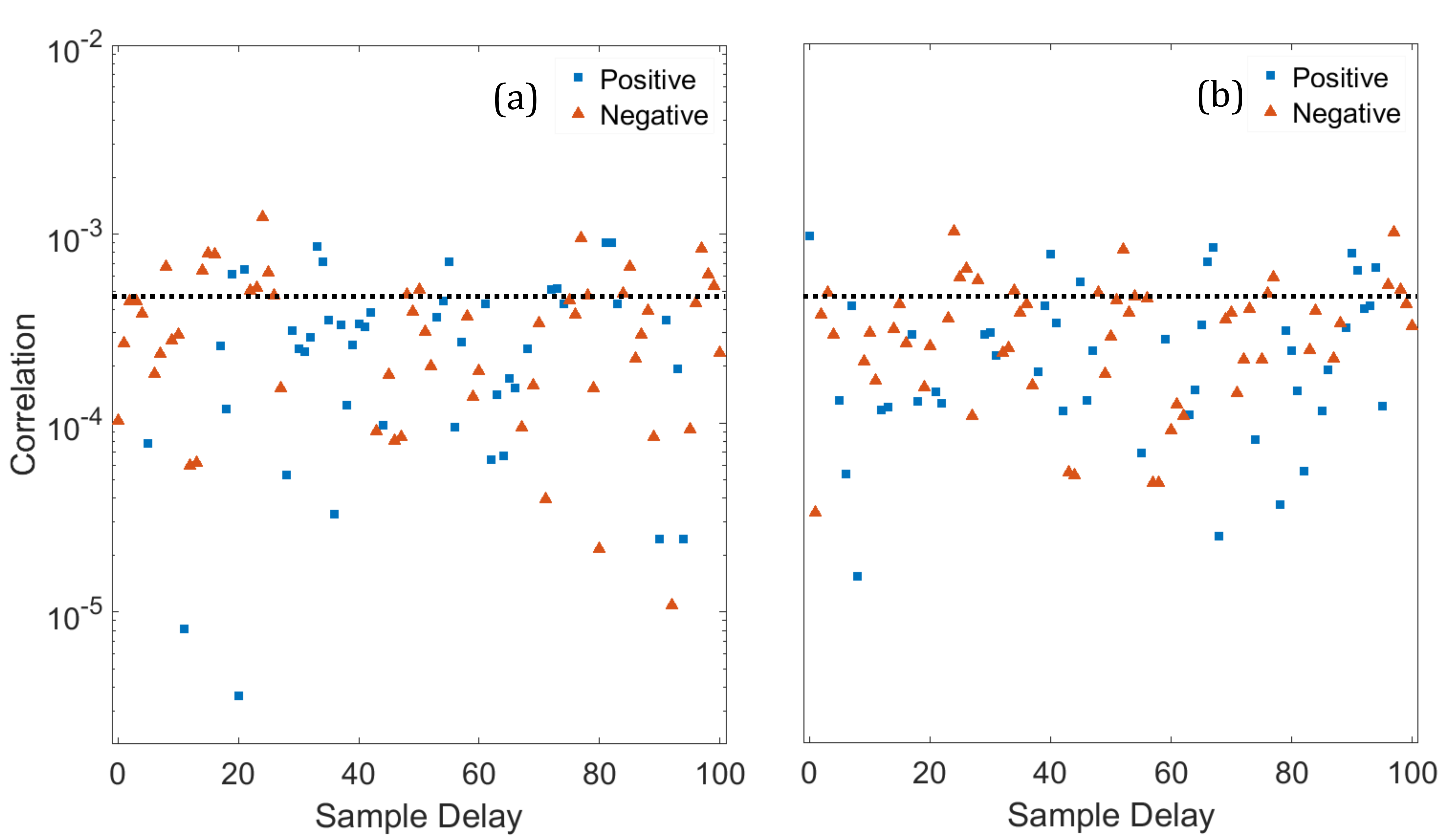}
\caption{Correlation between the output of a single two-source extractor and its inputs 1(a) and 2(b). Both positive (squares) and negative (triangles) correlations are plotted, with the y-axis shared between plots. The black dotted line represents the correlation of perfectly random data for the sample size, $4.8\times 10^{-4}$.}
\label{fig3}
\end{figure}
The two-source extractor we implement is strong given perfectly independent inputs i.e., the output is uncorrelated to either of the inputs.
 We quantify the effect of experimental imperfections in the input independence by calculating the cross-correlation between each input and the output as a measure of extractor strength, shown in Fig. 3 for $4.3\times 10^6$ samples. The theoretical correlation of perfectly random data for the sample size is $4.8\times 10^{-4}$.

\section{Conclusion}
In summary, we have demonstrated a high-speed parallel/ multiplexed quantum random number generator, a configuration ideally suited to a range of platforms, as well as capable of enhancing real time QRNG rates. Parallelisation of random number generation is an effective way to increase the real-time bit rate of QRNG's, and to supply quantum random numbers to distributed or cluster based computation and parallel communication systems. Furthermore, the parallel architecture allows us to demonstrate a high-speed un-keyed strong extraction(C) to create random numbers without the need for an external provider of uniform random seeds. True randomness sources that do not need a random seed have practical security by relying only on the validity of the partially random sources, and not requiring an external source of true randomness.

The highly specialized nature of the optical and electronic components makes a system on a single circuit board the realistic short-term integration option for GHz rate QRNG. To continue the scaling of this system to hundreds of channels full integration of a high power laser, waveguides, photodiodes, and processing electronics on a single chip will be necessary, and silicon offers a suitable platform for both electronic and optical components \cite{Silverstone:16}. The simplicity of the entropy source makes quantum vacuum fluctuations an excellent choice for parallelisation with Gbps single channel bitrates reported. 

\section*{Acknowledgements}
The authors thank Stefan Morley and Xingxing Xing for electronics support and Zachary Vernon for comments on the manuscript. BH is supported by the Australian Government Research Training Program Scholarship. This research is financially supported by the Australian Research Council Centre of Excellence for Quantum Computation and Communication Technology (CE170100012) and the Griffith University Research Infrastructure Programme.This work was performed in part at the Queensland node of the Australian National Fabrication Facility, a company established under the National Collaborative Research Infrastructure Strategy to provide nano- and microfabrication facilities for Australia's researchers. 


\begin{thebibliography}{44}%
	\makeatletter
	\providecommand \@ifxundefined [1]{%
		\@ifx{#1\undefined}
	}%
	\providecommand \@ifnum [1]{%
		\ifnum #1\expandafter \@firstoftwo
		\else \expandafter \@secondoftwo
		\fi
	}%
	\providecommand \@ifx [1]{%
		\ifx #1\expandafter \@firstoftwo
		\else \expandafter \@secondoftwo
		\fi
	}%
	\providecommand \natexlab [1]{#1}%
	\providecommand \enquote  [1]{``#1''}%
	\providecommand \bibnamefont  [1]{#1}%
	\providecommand \bibfnamefont [1]{#1}%
	\providecommand \citenamefont [1]{#1}%
	\providecommand \href@noop [0]{\@secondoftwo}%
	\providecommand \href [0]{\begingroup \@sanitize@url \@href}%
	\providecommand \@href[1]{\@@startlink{#1}\@@href}%
	\providecommand \@@href[1]{\endgroup#1\@@endlink}%
	\providecommand \@sanitize@url [0]{\catcode `\\12\catcode `\$12\catcode
		`\&12\catcode `\#12\catcode `\^12\catcode `\_12\catcode `\%12\relax}%
	\providecommand \@@startlink[1]{}%
	\providecommand \@@endlink[0]{}%
	\providecommand \url  [0]{\begingroup\@sanitize@url \@url }%
	\providecommand \@url [1]{\endgroup\@href {#1}{\urlprefix }}%
	\providecommand \urlprefix  [0]{URL }%
	\providecommand \Eprint [0]{\href }%
	\providecommand \doibase [0]{http://dx.doi.org/}%
	\providecommand \selectlanguage [0]{\@gobble}%
	\providecommand \bibinfo  [0]{\@secondoftwo}%
	\providecommand \bibfield  [0]{\@secondoftwo}%
	\providecommand \translation [1]{[#1]}%
	\providecommand \BibitemOpen [0]{}%
	\providecommand \bibitemStop [0]{}%
	\providecommand \bibitemNoStop [0]{.\EOS\space}%
	\providecommand \EOS [0]{\spacefactor3000\relax}%
	\providecommand \BibitemShut  [1]{\csname bibitem#1\endcsname}%
	\let\auto@bib@innerbib\@empty
	\bibitem [{\citenamefont {Ware}(1967)}]{Ware:67}%
	\BibitemOpen
	\bibfield  {author} {\bibinfo {author} {\bibfnamefont {W.~H.}\ \bibnamefont
			{Ware}},\ }in\ \href@noop {} {\emph {\bibinfo {booktitle} {Proceedings of the
				Spring Joint Computer Conference}}}\ (\bibinfo {year} {1967})\ pp.\ \bibinfo
	{pages} {279--282}\BibitemShut {NoStop}%
	\bibitem [{\citenamefont {Zhang}\ \emph {et~al.}(2014)\citenamefont {Zhang},
		\citenamefont {Aungskunsiri}, \citenamefont {Mart\'{\i}n-L\'{o}pez},
		\citenamefont {Wabnig}, \citenamefont {Lobino}, \citenamefont {Nock},
		\citenamefont {Munns}, \citenamefont {Bonneau}, \citenamefont {Jiang},
		\citenamefont {Li}, \citenamefont {Laing}, \citenamefont {Rarity},
		\citenamefont {Niskanen}, \citenamefont {Thompson},\ and\ \citenamefont
		{O'Brien}}]{Zhang:14}%
	\BibitemOpen
	\bibfield  {author} {\bibinfo {author} {\bibfnamefont {P.}~\bibnamefont
			{Zhang}}, \bibinfo {author} {\bibfnamefont {K.}~\bibnamefont {Aungskunsiri}},
		\bibinfo {author} {\bibfnamefont {E.}~\bibnamefont {Mart\'{\i}n-L\'{o}pez}},
		\bibinfo {author} {\bibfnamefont {J.}~\bibnamefont {Wabnig}}, \bibinfo
		{author} {\bibfnamefont {M.}~\bibnamefont {Lobino}}, \bibinfo {author}
		{\bibfnamefont {R.~W.}\ \bibnamefont {Nock}}, \bibinfo {author}
		{\bibfnamefont {J.}~\bibnamefont {Munns}}, \bibinfo {author} {\bibfnamefont
			{D.}~\bibnamefont {Bonneau}}, \bibinfo {author} {\bibfnamefont
			{P.}~\bibnamefont {Jiang}}, \bibinfo {author} {\bibfnamefont {H.~W.}\
			\bibnamefont {Li}}, \bibinfo {author} {\bibfnamefont {A.}~\bibnamefont
			{Laing}}, \bibinfo {author} {\bibfnamefont {J.~G.}\ \bibnamefont {Rarity}},
		\bibinfo {author} {\bibfnamefont {A.~O.}\ \bibnamefont {Niskanen}}, \bibinfo
		{author} {\bibfnamefont {M.~G.}\ \bibnamefont {Thompson}}, \ and\ \bibinfo
		{author} {\bibfnamefont {J.~L.}\ \bibnamefont {O'Brien}},\ }\href {\doibase
		10.1103/PhysRevLett.112.130501} {\bibfield  {journal} {\bibinfo  {journal}
			{Physical Review Letters}\ }\textbf {\bibinfo {volume} {112}},\ \bibinfo
		{pages} {130501} (\bibinfo {year} {2014})}\BibitemShut {NoStop}%
	\bibitem [{\citenamefont {Debian}()}]{Debian}%
	\BibitemOpen
	\bibfield  {author} {\bibinfo {author} {\bibnamefont {Debian}},\ }\href@noop
	{} {\emph {\bibinfo {title} {Debian - Security Information - DSA-1571-1
				openssl}}},\ \bibinfo {organization}
	{https://www.debian.org/security/2008/dsa-1571}\BibitemShut {NoStop}%
	\bibitem [{\citenamefont {Dorrendorf}\ \emph {et~al.}(2009)\citenamefont
		{Dorrendorf}, \citenamefont {Gutterman},\ and\ \citenamefont
		{Pinkas}}]{Dorrendorf:09}%
	\BibitemOpen
	\bibfield  {author} {\bibinfo {author} {\bibfnamefont {L.}~\bibnamefont
			{Dorrendorf}}, \bibinfo {author} {\bibfnamefont {Z.}~\bibnamefont
			{Gutterman}}, \ and\ \bibinfo {author} {\bibfnamefont {B.}~\bibnamefont
			{Pinkas}},\ }\href {\doibase 10.1145/1609956.1609966} {\bibfield  {journal}
		{\bibinfo  {journal} {ACM Transactions on Information and System Security}\
		}\textbf {\bibinfo {volume} {13}},\ \bibinfo {pages} {1} (\bibinfo {year}
		{2009})}\BibitemShut {NoStop}%
	\bibitem [{\citenamefont {Nohl}\ \emph {et~al.}(2008)\citenamefont {Nohl},
		\citenamefont {Evans}, \citenamefont {Starbug},\ and\ \citenamefont
		{Plotz}}]{Nohl:08}%
	\BibitemOpen
	\bibfield  {author} {\bibinfo {author} {\bibfnamefont {K.}~\bibnamefont
			{Nohl}}, \bibinfo {author} {\bibfnamefont {D.}~\bibnamefont {Evans}},
		\bibinfo {author} {\bibnamefont {Starbug}}, \ and\ \bibinfo {author}
		{\bibfnamefont {H.}~\bibnamefont {Plotz}},\ }in\ \href@noop {} {\emph
		{\bibinfo {booktitle} {17th USENIX Security Symposium}}}\ (\bibinfo {year}
	{2008})\ pp.\ \bibinfo {pages} {185--193}\BibitemShut {NoStop}%
	\bibitem [{\citenamefont {Herrero-Collantes}\ and\ \citenamefont
		{Garcia-Escartin}(2017)}]{H-C:17}%
	\BibitemOpen
	\bibfield  {author} {\bibinfo {author} {\bibfnamefont {M.}~\bibnamefont
			{Herrero-Collantes}}\ and\ \bibinfo {author} {\bibfnamefont {J.~C.}\
			\bibnamefont {Garcia-Escartin}},\ }\href {\doibase
		10.1103/RevModPhys.89.015004} {\bibfield  {journal} {\bibinfo  {journal}
			{Reviews of Modern Physics}\ }\textbf {\bibinfo {volume} {89}},\ \bibinfo
		{pages} {1} (\bibinfo {year} {2017})}\BibitemShut {NoStop}%
	\bibitem [{\citenamefont {Ma}\ \emph {et~al.}(2016)\citenamefont {Ma},
		\citenamefont {Yuan}, \citenamefont {Qi},\ and\ \citenamefont
		{Zhang}}]{Ma:16}%
	\BibitemOpen
	\bibfield  {author} {\bibinfo {author} {\bibfnamefont {X.}~\bibnamefont
			{Ma}}, \bibinfo {author} {\bibfnamefont {X.}~\bibnamefont {Yuan}}, \bibinfo
		{author} {\bibfnamefont {B.}~\bibnamefont {Qi}}, \ and\ \bibinfo {author}
		{\bibfnamefont {Z.}~\bibnamefont {Zhang}},\ }\href {\doibase
		10.1038/npjqi.2016.21} {\bibfield  {journal} {\bibinfo  {journal} {npj
				Quantum Information}\ }\textbf {\bibinfo {volume} {221}},\ \bibinfo {pages}
		{1} (\bibinfo {year} {2016})}\BibitemShut {NoStop}%
	\bibitem [{\citenamefont {Hart}\ \emph {et~al.}(2017)\citenamefont {Hart},
		\citenamefont {Terashima}, \citenamefont {Uchida}, \citenamefont
		{Baumgartner}, \citenamefont {Murphy},\ and\ \citenamefont {Roy}}]{Hart:17}%
	\BibitemOpen
	\bibfield  {author} {\bibinfo {author} {\bibfnamefont {J.~D.}\ \bibnamefont
			{Hart}}, \bibinfo {author} {\bibfnamefont {Y.}~\bibnamefont {Terashima}},
		\bibinfo {author} {\bibfnamefont {A.}~\bibnamefont {Uchida}}, \bibinfo
		{author} {\bibfnamefont {G.~B.}\ \bibnamefont {Baumgartner}}, \bibinfo
		{author} {\bibfnamefont {T.~E.}\ \bibnamefont {Murphy}}, \ and\ \bibinfo
		{author} {\bibfnamefont {R.}~\bibnamefont {Roy}},\ }\href {\doibase
		10.1063/1.5000056} {\bibfield  {journal} {\bibinfo  {journal} {APL
				Photonics}\ }\textbf {\bibinfo {volume} {2}},\ \bibinfo {pages} {1} (\bibinfo
		{year} {2017})}\BibitemShut {NoStop}%
	\bibitem [{\citenamefont {Nie}\ \emph {et~al.}(2016)\citenamefont {Nie},
		\citenamefont {Guan}, \citenamefont {Zhou}, \citenamefont {Zhang},
		\citenamefont {Ma},\ and\ \citenamefont {Pan}}]{Nie:16}%
	\BibitemOpen
	\bibfield  {author} {\bibinfo {author} {\bibfnamefont {Y.~Q.}\ \bibnamefont
			{Nie}}, \bibinfo {author} {\bibfnamefont {J.~Y.}\ \bibnamefont {Guan}},
		\bibinfo {author} {\bibfnamefont {H.}~\bibnamefont {Zhou}}, \bibinfo {author}
		{\bibfnamefont {Q.}~\bibnamefont {Zhang}}, \bibinfo {author} {\bibfnamefont
			{X.}~\bibnamefont {Ma}}, \ and\ \bibinfo {author} {\bibfnamefont {J.~W.}\
			\bibnamefont {Pan}},\ }\href {\doibase 10.1103/PhysRevA.94.060301} {\bibfield
		{journal} {\bibinfo  {journal} {Physical Review A}\ }\textbf {\bibinfo
			{volume} {94}},\ \bibinfo {pages} {060301} (\bibinfo {year}
		{2016})}\BibitemShut {NoStop}%
	\bibitem [{\citenamefont {Pironio}\ \emph {et~al.}(2010)\citenamefont
		{Pironio}, \citenamefont {Acin}, \citenamefont {Massar}, \citenamefont {Boyer
			de~la Giroday}, \citenamefont {Matsukevich}, \citenamefont {Maunz},
		\citenamefont {Olmschenk}, \citenamefont {Hayes}, \citenamefont {Luo},
		\citenamefont {Manning},\ and\ \citenamefont {Monroe}}]{Pironio:10}%
	\BibitemOpen
	\bibfield  {author} {\bibinfo {author} {\bibfnamefont {S.}~\bibnamefont
			{Pironio}}, \bibinfo {author} {\bibfnamefont {A.}~\bibnamefont {Acin}},
		\bibinfo {author} {\bibfnamefont {S.}~\bibnamefont {Massar}}, \bibinfo
		{author} {\bibfnamefont {A.}~\bibnamefont {Boyer de~la Giroday}}, \bibinfo
		{author} {\bibfnamefont {D.~N.}\ \bibnamefont {Matsukevich}}, \bibinfo
		{author} {\bibfnamefont {P.}~\bibnamefont {Maunz}}, \bibinfo {author}
		{\bibfnamefont {S.}~\bibnamefont {Olmschenk}}, \bibinfo {author}
		{\bibfnamefont {D.}~\bibnamefont {Hayes}}, \bibinfo {author} {\bibfnamefont
			{L.}~\bibnamefont {Luo}}, \bibinfo {author} {\bibfnamefont {T.~A.}\
			\bibnamefont {Manning}}, \ and\ \bibinfo {author} {\bibfnamefont
			{C.}~\bibnamefont {Monroe}},\ }\href {\doibase 10.1038/nature09008}
	{\bibfield  {journal} {\bibinfo  {journal} {Nature}\ }\textbf {\bibinfo
			{volume} {464}},\ \bibinfo {pages} {1021} (\bibinfo {year}
		{2010})}\BibitemShut {NoStop}%
	\bibitem [{\citenamefont {Bierhorst}\ \emph {et~al.}(2017)\citenamefont
		{Bierhorst}, \citenamefont {Knill}, \citenamefont {Glancy}, \citenamefont
		{Mink}, \citenamefont {Jordan}, \citenamefont {Rommal}, \citenamefont {Liu},
		\citenamefont {Christensen}, \citenamefont {Nam},\ and\ \citenamefont
		{Shalm}}]{Bierhorst}%
	\BibitemOpen
	\bibfield  {author} {\bibinfo {author} {\bibfnamefont {P.}~\bibnamefont
			{Bierhorst}}, \bibinfo {author} {\bibfnamefont {E.}~\bibnamefont {Knill}},
		\bibinfo {author} {\bibfnamefont {S.}~\bibnamefont {Glancy}}, \bibinfo
		{author} {\bibfnamefont {A.}~\bibnamefont {Mink}}, \bibinfo {author}
		{\bibfnamefont {S.}~\bibnamefont {Jordan}}, \bibinfo {author} {\bibfnamefont
			{A.}~\bibnamefont {Rommal}}, \bibinfo {author} {\bibfnamefont {A.~K.}\
			\bibnamefont {Liu}}, \bibinfo {author} {\bibfnamefont {B.}~\bibnamefont
			{Christensen}}, \bibinfo {author} {\bibfnamefont {S.~W.}\ \bibnamefont
			{Nam}}, \ and\ \bibinfo {author} {\bibfnamefont {L.~K.}\ \bibnamefont
			{Shalm}},\ }\href@noop {} {\  (\bibinfo {year} {2017})},\ \bibinfo {note}
	{arXiv 1702.05178}\BibitemShut {NoStop}%
	\bibitem [{\citenamefont {Kulikov}\ \emph {et~al.}(2017)\citenamefont
		{Kulikov}, \citenamefont {Jerger}, \citenamefont {Poto\u{c}nik},
		\citenamefont {Wallraff},\ and\ \citenamefont {Fedorov}}]{Kulikov:17}%
	\BibitemOpen
	\bibfield  {author} {\bibinfo {author} {\bibfnamefont {A.}~\bibnamefont
			{Kulikov}}, \bibinfo {author} {\bibfnamefont {M.}~\bibnamefont {Jerger}},
		\bibinfo {author} {\bibfnamefont {A.}~\bibnamefont {Poto\u{c}nik}}, \bibinfo
		{author} {\bibfnamefont {A.}~\bibnamefont {Wallraff}}, \ and\ \bibinfo
		{author} {\bibfnamefont {A.}~\bibnamefont {Fedorov}},\ }\href {\doibase
		10.1103/PhysRevLett.119.240501} {\bibfield  {journal} {\bibinfo  {journal}
			{Physical Review Letters}\ }\textbf {\bibinfo {volume} {119}},\ \bibinfo
		{pages} {240501} (\bibinfo {year} {2017})}\BibitemShut {NoStop}%
	\bibitem [{\citenamefont {Abbott}\ \emph {et~al.}(2012)\citenamefont {Abbott},
		\citenamefont {Calude}, \citenamefont {Conder},\ and\ \citenamefont
		{Svozil}}]{Abbott:12}%
	\BibitemOpen
	\bibfield  {author} {\bibinfo {author} {\bibfnamefont {A.~A.}\ \bibnamefont
			{Abbott}}, \bibinfo {author} {\bibfnamefont {C.~S.}\ \bibnamefont {Calude}},
		\bibinfo {author} {\bibfnamefont {J.}~\bibnamefont {Conder}}, \ and\ \bibinfo
		{author} {\bibfnamefont {K.}~\bibnamefont {Svozil}},\ }\href {\doibase
		10.1103/PhysRevA.86.062109} {\bibfield  {journal} {\bibinfo  {journal}
			{Physical Review A}\ }\textbf {\bibinfo {volume} {86}},\ \bibinfo {pages}
		{062109} (\bibinfo {year} {2012})}\BibitemShut {NoStop}%
	\bibitem [{\citenamefont {Liu}\ \emph {et~al.}(2018)\citenamefont {Liu},
		\citenamefont {Yuan}, \citenamefont {Li}, \citenamefont {Zhang},
		\citenamefont {Zhao}, \citenamefont {Zhong}, \citenamefont {Cao},
		\citenamefont {Li}, \citenamefont {Chen}, \citenamefont {Li}, \citenamefont
		{Peng}, \citenamefont {Chen}, \citenamefont {Peng}, \citenamefont {Shi},
		\citenamefont {Wang}, \citenamefont {You}, \citenamefont {Ma}, \citenamefont
		{Fan}, \citenamefont {Zhang},\ and\ \citenamefont {Pan}}]{Liu:18}%
	\BibitemOpen
	\bibfield  {author} {\bibinfo {author} {\bibfnamefont {Y.}~\bibnamefont
			{Liu}}, \bibinfo {author} {\bibfnamefont {X.}~\bibnamefont {Yuan}}, \bibinfo
		{author} {\bibfnamefont {M.-H.}\ \bibnamefont {Li}}, \bibinfo {author}
		{\bibfnamefont {W.}~\bibnamefont {Zhang}}, \bibinfo {author} {\bibfnamefont
			{Q.}~\bibnamefont {Zhao}}, \bibinfo {author} {\bibfnamefont {J.}~\bibnamefont
			{Zhong}}, \bibinfo {author} {\bibfnamefont {Y.}~\bibnamefont {Cao}}, \bibinfo
		{author} {\bibfnamefont {Y.-H.}\ \bibnamefont {Li}}, \bibinfo {author}
		{\bibfnamefont {L.-K.}\ \bibnamefont {Chen}}, \bibinfo {author}
		{\bibfnamefont {H.}~\bibnamefont {Li}}, \bibinfo {author} {\bibfnamefont
			{T.}~\bibnamefont {Peng}}, \bibinfo {author} {\bibfnamefont {Y.-A.}\
			\bibnamefont {Chen}}, \bibinfo {author} {\bibfnamefont {C.-Z.}\ \bibnamefont
			{Peng}}, \bibinfo {author} {\bibfnamefont {S.-C.}\ \bibnamefont {Shi}},
		\bibinfo {author} {\bibfnamefont {Z.}~\bibnamefont {Wang}}, \bibinfo {author}
		{\bibfnamefont {L.}~\bibnamefont {You}}, \bibinfo {author} {\bibfnamefont
			{X.}~\bibnamefont {Ma}}, \bibinfo {author} {\bibfnamefont {J.}~\bibnamefont
			{Fan}}, \bibinfo {author} {\bibfnamefont {Q.}~\bibnamefont {Zhang}}, \ and\
		\bibinfo {author} {\bibfnamefont {J.-W.}\ \bibnamefont {Pan}},\ }\href
	{\doibase 10.1103/PhysRevLett.120.010503} {\bibfield  {journal} {\bibinfo
			{journal} {Physical Review Letters}\ }\textbf {\bibinfo {volume} {120}},\
		\bibinfo {pages} {010503} (\bibinfo {year} {2018})}\BibitemShut {NoStop}%
	\bibitem [{\citenamefont {Gabriel}\ \emph {et~al.}(2010)\citenamefont
		{Gabriel}, \citenamefont {Wittmann}, \citenamefont {Sych}, \citenamefont
		{Dong}, \citenamefont {Mauerer}, \citenamefont {Andersen}, \citenamefont
		{Marquadt},\ and\ \citenamefont {Leuchs}}]{Gabriel:10}%
	\BibitemOpen
	\bibfield  {author} {\bibinfo {author} {\bibfnamefont {C.}~\bibnamefont
			{Gabriel}}, \bibinfo {author} {\bibfnamefont {C.}~\bibnamefont {Wittmann}},
		\bibinfo {author} {\bibfnamefont {D.}~\bibnamefont {Sych}}, \bibinfo {author}
		{\bibfnamefont {R.}~\bibnamefont {Dong}}, \bibinfo {author} {\bibfnamefont
			{W.}~\bibnamefont {Mauerer}}, \bibinfo {author} {\bibfnamefont {U.~L.}\
			\bibnamefont {Andersen}}, \bibinfo {author} {\bibfnamefont {C.}~\bibnamefont
			{Marquadt}}, \ and\ \bibinfo {author} {\bibfnamefont {G.}~\bibnamefont
			{Leuchs}},\ }\href {\doibase 10.1038/nphoton.2010.197} {\bibfield  {journal}
		{\bibinfo  {journal} {Nature Photonics}\ }\textbf {\bibinfo {volume} {4}},\
		\bibinfo {pages} {711} (\bibinfo {year} {2010})}\BibitemShut {NoStop}%
	\bibitem [{\citenamefont {Li}\ \emph {et~al.}(2014)\citenamefont {Li},
		\citenamefont {Wang}, \citenamefont {Li}, \citenamefont {Xu}, \citenamefont
		{Wang},\ and\ \citenamefont {Wang}}]{Li:14}%
	\BibitemOpen
	\bibfield  {author} {\bibinfo {author} {\bibfnamefont {L.}~\bibnamefont
			{Li}}, \bibinfo {author} {\bibfnamefont {A.}~\bibnamefont {Wang}}, \bibinfo
		{author} {\bibfnamefont {P.}~\bibnamefont {Li}}, \bibinfo {author}
		{\bibfnamefont {H.}~\bibnamefont {Xu}}, \bibinfo {author} {\bibfnamefont
			{L.}~\bibnamefont {Wang}}, \ and\ \bibinfo {author} {\bibfnamefont
			{Y.}~\bibnamefont {Wang}},\ }\href {\doibase 10.1109/JPHOT.2014.2304555}
	{\bibfield  {journal} {\bibinfo  {journal} {IEEE Photonics Journal}\ }\textbf
		{\bibinfo {volume} {6}} (\bibinfo {year} {2014}),\
		10.1109/JPHOT.2014.2304555}\BibitemShut {NoStop}%
	\bibitem [{\citenamefont {Lunghi}\ \emph {et~al.}(2015)\citenamefont {Lunghi},
		\citenamefont {Brask}, \citenamefont {Lim}, \citenamefont {Lavigne},
		\citenamefont {Bowles}, \citenamefont {Martin}, \citenamefont {Zbinden},\
		and\ \citenamefont {Brunner}}]{Lunghi:15}%
	\BibitemOpen
	\bibfield  {author} {\bibinfo {author} {\bibfnamefont {T.}~\bibnamefont
			{Lunghi}}, \bibinfo {author} {\bibfnamefont {J.~B.}\ \bibnamefont {Brask}},
		\bibinfo {author} {\bibfnamefont {C.~C.~W.}\ \bibnamefont {Lim}}, \bibinfo
		{author} {\bibfnamefont {Q.}~\bibnamefont {Lavigne}}, \bibinfo {author}
		{\bibfnamefont {J.}~\bibnamefont {Bowles}}, \bibinfo {author} {\bibfnamefont
			{A.}~\bibnamefont {Martin}}, \bibinfo {author} {\bibfnamefont
			{H.}~\bibnamefont {Zbinden}}, \ and\ \bibinfo {author} {\bibfnamefont
			{N.}~\bibnamefont {Brunner}},\ }\href {\doibase
		10.1103/PhysRevLett.114.150501} {\bibfield  {journal} {\bibinfo  {journal}
			{Physical Review Letters}\ }\textbf {\bibinfo {volume} {114}},\ \bibinfo
		{pages} {150501} (\bibinfo {year} {2015})}\BibitemShut {NoStop}%
	\bibitem [{\citenamefont {Mitchell}\ \emph {et~al.}(2015)\citenamefont
		{Mitchell}, \citenamefont {Abellan},\ and\ \citenamefont
		{Amaya}}]{Mitchell:15}%
	\BibitemOpen
	\bibfield  {author} {\bibinfo {author} {\bibfnamefont {M.~W.}\ \bibnamefont
			{Mitchell}}, \bibinfo {author} {\bibfnamefont {C.}~\bibnamefont {Abellan}}, \
		and\ \bibinfo {author} {\bibfnamefont {W.}~\bibnamefont {Amaya}},\ }\href
	{\doibase 10.1103/PhysRevA.91.012314} {\bibfield  {journal} {\bibinfo
			{journal} {Physical Review A}\ }\textbf {\bibinfo {volume} {91}},\ \bibinfo
		{pages} {012314} (\bibinfo {year} {2015})}\BibitemShut {NoStop}%
	\bibitem [{\citenamefont {Marangon}\ \emph {et~al.}(2017)\citenamefont
		{Marangon}, \citenamefont {Vallone},\ and\ \citenamefont
		{Villoresi}}]{Marangon:2017}%
	\BibitemOpen
	\bibfield  {author} {\bibinfo {author} {\bibfnamefont {D.~G.}\ \bibnamefont
			{Marangon}}, \bibinfo {author} {\bibfnamefont {G.}~\bibnamefont {Vallone}}, \
		and\ \bibinfo {author} {\bibfnamefont {P.}~\bibnamefont {Villoresi}},\ }\href
	{\doibase 10.1103/PhysRevLett.118.060503} {\bibfield  {journal} {\bibinfo
			{journal} {Physical Review Letters}\ }\textbf {\bibinfo {volume} {118}},\
		\bibinfo {pages} {060503} (\bibinfo {year} {2017})}\BibitemShut {NoStop}%
	\bibitem [{\citenamefont {Virte}\ \emph {et~al.}(2014)\citenamefont {Virte},
		\citenamefont {Mercier}, \citenamefont {Thienpont}, \citenamefont
		{Panajotov},\ and\ \citenamefont {Sciamanna}}]{Virte:14}%
	\BibitemOpen
	\bibfield  {author} {\bibinfo {author} {\bibfnamefont {M.}~\bibnamefont
			{Virte}}, \bibinfo {author} {\bibfnamefont {E.}~\bibnamefont {Mercier}},
		\bibinfo {author} {\bibfnamefont {H.}~\bibnamefont {Thienpont}}, \bibinfo
		{author} {\bibfnamefont {K.}~\bibnamefont {Panajotov}}, \ and\ \bibinfo
		{author} {\bibfnamefont {M.}~\bibnamefont {Sciamanna}},\ }\href {\doibase
		10.1364/OE.22.017271} {\bibfield  {journal} {\bibinfo  {journal} {Optics
				Express}\ }\textbf {\bibinfo {volume} {22}},\ \bibinfo {pages} {17271}
		(\bibinfo {year} {2014})}\BibitemShut {NoStop}%
	\bibitem [{\citenamefont {Wayne}\ \emph {et~al.}(2009)\citenamefont {Wayne},
		\citenamefont {Jeffrey}, \citenamefont {Akselrod},\ and\ \citenamefont
		{Kwiat}}]{Wayne:09}%
	\BibitemOpen
	\bibfield  {author} {\bibinfo {author} {\bibfnamefont {M.~A.}\ \bibnamefont
			{Wayne}}, \bibinfo {author} {\bibfnamefont {E.~R.}\ \bibnamefont {Jeffrey}},
		\bibinfo {author} {\bibfnamefont {G.~M.}\ \bibnamefont {Akselrod}}, \ and\
		\bibinfo {author} {\bibfnamefont {P.~G.}\ \bibnamefont {Kwiat}},\ }\href
	{\doibase 10.1080/09500340802553244} {\bibfield  {journal} {\bibinfo
			{journal} {Journal of Modern Optics}\ }\textbf {\bibinfo {volume} {56}},\
		\bibinfo {pages} {516} (\bibinfo {year} {2009})}\BibitemShut {NoStop}%
	\bibitem [{\citenamefont {Xu}\ \emph {et~al.}(2012)\citenamefont {Xu},
		\citenamefont {Qi}, \citenamefont {Ma}, \citenamefont {Zheng},\ and\
		\citenamefont {Lo}}]{Xu:12}%
	\BibitemOpen
	\bibfield  {author} {\bibinfo {author} {\bibfnamefont {F.}~\bibnamefont
			{Xu}}, \bibinfo {author} {\bibfnamefont {B.}~\bibnamefont {Qi}}, \bibinfo
		{author} {\bibfnamefont {X.}~\bibnamefont {Ma}}, \bibinfo {author}
		{\bibfnamefont {H.}~\bibnamefont {Zheng}}, \ and\ \bibinfo {author}
		{\bibfnamefont {H.~K.}\ \bibnamefont {Lo}},\ }\href {\doibase
		10.1364/OE.20.012366} {\bibfield  {journal} {\bibinfo  {journal} {Optics
				Express}\ }\textbf {\bibinfo {volume} {20}},\ \bibinfo {pages} {12366}
		(\bibinfo {year} {2012})}\BibitemShut {NoStop}%
	\bibitem [{\citenamefont {Sakuraba}\ \emph {et~al.}(2015)\citenamefont
		{Sakuraba}, \citenamefont {Iwakawa}, \citenamefont {Kanno},\ and\
		\citenamefont {Uchida}}]{Sakuraba:15}%
	\BibitemOpen
	\bibfield  {author} {\bibinfo {author} {\bibfnamefont {R.}~\bibnamefont
			{Sakuraba}}, \bibinfo {author} {\bibfnamefont {K.}~\bibnamefont {Iwakawa}},
		\bibinfo {author} {\bibfnamefont {K.}~\bibnamefont {Kanno}}, \ and\ \bibinfo
		{author} {\bibfnamefont {A.}~\bibnamefont {Uchida}},\ }\href {\doibase
		10.1364/OE.23.001470} {\bibfield  {journal} {\bibinfo  {journal} {Optics
				Express}\ }\textbf {\bibinfo {volume} {23}},\ \bibinfo {pages} {1470}
		(\bibinfo {year} {2015})}\BibitemShut {NoStop}%
	\bibitem [{\citenamefont {Gr\"{a}fe}\ \emph {et~al.}(2014)\citenamefont
		{Gr\"{a}fe}, \citenamefont {Heilmann}, \citenamefont {Perez-Leija},
		\citenamefont {Keil}, \citenamefont {Dreisow}, \citenamefont {Heinrich},
		\citenamefont {Moya-Cessa}, \citenamefont {Nolte}, \citenamefont
		{Christodoulides},\ and\ \citenamefont {Szameit}}]{Grafe:14}%
	\BibitemOpen
	\bibfield  {author} {\bibinfo {author} {\bibfnamefont {M.}~\bibnamefont
			{Gr\"{a}fe}}, \bibinfo {author} {\bibfnamefont {R.}~\bibnamefont {Heilmann}},
		\bibinfo {author} {\bibfnamefont {A.}~\bibnamefont {Perez-Leija}}, \bibinfo
		{author} {\bibfnamefont {R.}~\bibnamefont {Keil}}, \bibinfo {author}
		{\bibfnamefont {F.}~\bibnamefont {Dreisow}}, \bibinfo {author} {\bibfnamefont
			{M.}~\bibnamefont {Heinrich}}, \bibinfo {author} {\bibfnamefont
			{H.}~\bibnamefont {Moya-Cessa}}, \bibinfo {author} {\bibfnamefont
			{S.}~\bibnamefont {Nolte}}, \bibinfo {author} {\bibfnamefont {D.~N.}\
			\bibnamefont {Christodoulides}}, \ and\ \bibinfo {author} {\bibfnamefont
			{A.}~\bibnamefont {Szameit}},\ }\href {\doibase 10.1038/nphoton.2014.204}
	{\bibfield  {journal} {\bibinfo  {journal} {Nature Photonics}\ }\textbf
		{\bibinfo {volume} {8}},\ \bibinfo {pages} {791} (\bibinfo {year}
		{2014})}\BibitemShut {NoStop}%
	\bibitem [{\citenamefont {Haw}\ \emph {et~al.}(2015)\citenamefont {Haw},
		\citenamefont {Assad}, \citenamefont {Lance}, \citenamefont {Ng},
		\citenamefont {Sharma}, \citenamefont {Lam},\ and\ \citenamefont
		{Symul}}]{Haw:15}%
	\BibitemOpen
	\bibfield  {author} {\bibinfo {author} {\bibfnamefont {J.~Y.}\ \bibnamefont
			{Haw}}, \bibinfo {author} {\bibfnamefont {S.~M.}\ \bibnamefont {Assad}},
		\bibinfo {author} {\bibfnamefont {A.~M.}\ \bibnamefont {Lance}}, \bibinfo
		{author} {\bibfnamefont {N.~H.~Y.}\ \bibnamefont {Ng}}, \bibinfo {author}
		{\bibfnamefont {V.}~\bibnamefont {Sharma}}, \bibinfo {author} {\bibfnamefont
			{P.~K.}\ \bibnamefont {Lam}}, \ and\ \bibinfo {author} {\bibfnamefont
			{T.}~\bibnamefont {Symul}},\ }\href {\doibase
		10.1103/PhysRevApplied.3.054004} {\bibfield  {journal} {\bibinfo  {journal}
			{Physical Review Applied}\ }\textbf {\bibinfo {volume} {3}},\ \bibinfo
		{pages} {054004} (\bibinfo {year} {2015})}\BibitemShut {NoStop}%
	\bibitem [{\citenamefont {IDQuantique}(2016)}]{IDQuantique}%
	\BibitemOpen
	\bibfield  {author} {\bibinfo {author} {\bibnamefont {IDQuantique}},\
	}\href@noop {} {\emph {\bibinfo {title} {Quantis QRNG Brochure}}},\ \bibinfo
	{organization} {https://www.idquantique.com} (\bibinfo {year}
	{2016})\BibitemShut {NoStop}%
	\bibitem [{\citenamefont {Shaltiel}(2011)}]{Shaltiel:11}%
	\BibitemOpen
	\bibfield  {author} {\bibinfo {author} {\bibfnamefont {R.}~\bibnamefont
			{Shaltiel}},\ }in\ \href {\doibase 10.1007/978-3-642-22012-8_2} {\emph
		{\bibinfo {booktitle} {Lecture Notes in Computer Science}}},\ Vol.\ \bibinfo
	{volume} {6756}\ (\bibinfo {year} {2011})\ pp.\ \bibinfo {pages}
	{21--41}\BibitemShut {NoStop}%
	\bibitem [{\citenamefont {Kasher}\ and\ \citenamefont {Kempe}()}]{Kasher:10}%
	\BibitemOpen
	\bibfield  {author} {\bibinfo {author} {\bibfnamefont {R.}~\bibnamefont
			{Kasher}}\ and\ \bibinfo {author} {\bibfnamefont {J.}~\bibnamefont {Kempe}},\
	}in\ \href {\doibase 10.1007/978-3-642-15369-3_49} {\emph {\bibinfo
		{booktitle} {Approximation, Randomization, and Combinatorial Optimization.
			Algorithms and Techniques}}},\ \bibinfo {series} {Lecture Notes in Computer
	Science}, Vol.\ \bibinfo {volume} {6302},\ pp.\ \bibinfo {pages}
{656--669}\BibitemShut {NoStop}%
\bibitem [{\citenamefont {Ugajin}\ \emph {et~al.}(2017)\citenamefont {Ugajin},
	\citenamefont {Terashima}, \citenamefont {Iwakawa}, \citenamefont {Uchida},
	\citenamefont {Harayama}, \citenamefont {Yoshimura},\ and\ \citenamefont
	{Inubushi}}]{Ugajin:17}%
\BibitemOpen
\bibfield  {author} {\bibinfo {author} {\bibfnamefont {K.}~\bibnamefont
		{Ugajin}}, \bibinfo {author} {\bibfnamefont {Y.}~\bibnamefont {Terashima}},
	\bibinfo {author} {\bibfnamefont {K.}~\bibnamefont {Iwakawa}}, \bibinfo
	{author} {\bibfnamefont {A.}~\bibnamefont {Uchida}}, \bibinfo {author}
	{\bibfnamefont {T.}~\bibnamefont {Harayama}}, \bibinfo {author}
	{\bibfnamefont {K.}~\bibnamefont {Yoshimura}}, \ and\ \bibinfo {author}
	{\bibfnamefont {M.}~\bibnamefont {Inubushi}},\ }\href@noop {} {\bibfield
	{journal} {\bibinfo  {journal} {Optics Express}\ }\textbf {\bibinfo {volume}
		{25}},\ \bibinfo {pages} {6511} (\bibinfo {year} {2017})}\BibitemShut
{NoStop}%
\bibitem [{\citenamefont {Zhang}\ \emph {et~al.}(2016)\citenamefont {Zhang},
	\citenamefont {Nie}, \citenamefont {Zhou}, \citenamefont {Liang},
	\citenamefont {Ma}, \citenamefont {Zhang},\ and\ \citenamefont
	{Pan}}]{Zhang:16}%
\BibitemOpen
\bibfield  {author} {\bibinfo {author} {\bibfnamefont {X.-G.}\ \bibnamefont
		{Zhang}}, \bibinfo {author} {\bibfnamefont {Y.-Q.}\ \bibnamefont {Nie}},
	\bibinfo {author} {\bibfnamefont {H.}~\bibnamefont {Zhou}}, \bibinfo {author}
	{\bibfnamefont {H.}~\bibnamefont {Liang}}, \bibinfo {author} {\bibfnamefont
		{X.}~\bibnamefont {Ma}}, \bibinfo {author} {\bibfnamefont {J.}~\bibnamefont
		{Zhang}}, \ and\ \bibinfo {author} {\bibfnamefont {J.-W.}\ \bibnamefont
		{Pan}},\ }\href {\doibase 10.1063/1.4958663} {\bibfield  {journal} {\bibinfo
		{journal} {Review of Scientific Instruments}\ }\textbf {\bibinfo {volume}
		{87}},\ \bibinfo {pages} {076102} (\bibinfo {year} {2016})}\BibitemShut
{NoStop}%
\bibitem [{\citenamefont {Marangon}\ \emph {et~al.}(2018)\citenamefont
	{Marangon}, \citenamefont {Plews}, \citenamefont {Lucamarini}, \citenamefont
	{Dynes}, \citenamefont {Sharpe}, \citenamefont {Yuan},\ and\ \citenamefont
	{Shields}}]{Marangon:18}%
\BibitemOpen
\bibfield  {author} {\bibinfo {author} {\bibfnamefont {D.~G.}\ \bibnamefont
		{Marangon}}, \bibinfo {author} {\bibfnamefont {A.}~\bibnamefont {Plews}},
	\bibinfo {author} {\bibfnamefont {M.}~\bibnamefont {Lucamarini}}, \bibinfo
	{author} {\bibfnamefont {J.~F.}\ \bibnamefont {Dynes}}, \bibinfo {author}
	{\bibfnamefont {A.~W.}\ \bibnamefont {Sharpe}}, \bibinfo {author}
	{\bibfnamefont {Z.}~\bibnamefont {Yuan}}, \ and\ \bibinfo {author}
	{\bibfnamefont {A.~J.}\ \bibnamefont {Shields}},\ }\href@noop {} {\bibfield
	{journal} {\bibinfo  {journal} {Journal of Lightwave Technology}\ }\textbf
	{\bibinfo {volume} {36}},\ \bibinfo {pages} {3778} (\bibinfo {year}
	{2018})}\BibitemShut {NoStop}%
\bibitem [{\citenamefont {Shen}\ \emph {et~al.}(2010)\citenamefont {Shen},
	\citenamefont {Tian},\ and\ \citenamefont {Zou}}]{Shen:10}%
\BibitemOpen
\bibfield  {author} {\bibinfo {author} {\bibfnamefont {Y.}~\bibnamefont
		{Shen}}, \bibinfo {author} {\bibfnamefont {L.}~\bibnamefont {Tian}}, \ and\
	\bibinfo {author} {\bibfnamefont {H.}~\bibnamefont {Zou}},\ }\href {\doibase
	10.1103/PhysRevA.81.063814} {\bibfield  {journal} {\bibinfo  {journal}
		{Physical Review A}\ }\textbf {\bibinfo {volume} {81}},\ \bibinfo {pages}
	{063814} (\bibinfo {year} {2010})}\BibitemShut {NoStop}%
\bibitem [{\citenamefont {Kumar}\ \emph {et~al.}(2012)\citenamefont {Kumar},
	\citenamefont {Barrios}, \citenamefont {MacRae}, \citenamefont {Cairns},
	\citenamefont {Huntington},\ and\ \citenamefont {Lvovsky}}]{Kumar:12}%
\BibitemOpen
\bibfield  {author} {\bibinfo {author} {\bibfnamefont {R.}~\bibnamefont
		{Kumar}}, \bibinfo {author} {\bibfnamefont {E.}~\bibnamefont {Barrios}},
	\bibinfo {author} {\bibfnamefont {A.}~\bibnamefont {MacRae}}, \bibinfo
	{author} {\bibfnamefont {E.}~\bibnamefont {Cairns}}, \bibinfo {author}
	{\bibfnamefont {E.~H.}\ \bibnamefont {Huntington}}, \ and\ \bibinfo {author}
	{\bibfnamefont {A.~I.}\ \bibnamefont {Lvovsky}},\ }\href {\doibase
	10.1016/j.optcom.2012.07.103} {\bibfield  {journal} {\bibinfo  {journal}
		{Optics Communications}\ }\textbf {\bibinfo {volume} {285}},\ \bibinfo
	{pages} {24} (\bibinfo {year} {2012})}\BibitemShut {NoStop}%
\bibitem [{\citenamefont {Lenzini}\ \emph {et~al.}(2015)\citenamefont
	{Lenzini}, \citenamefont {Kasture}, \citenamefont {Haylock},\ and\
	\citenamefont {Lobino}}]{Lenzini:15}%
\BibitemOpen
\bibfield  {author} {\bibinfo {author} {\bibfnamefont {F.}~\bibnamefont
		{Lenzini}}, \bibinfo {author} {\bibfnamefont {S.}~\bibnamefont {Kasture}},
	\bibinfo {author} {\bibfnamefont {B.}~\bibnamefont {Haylock}}, \ and\
	\bibinfo {author} {\bibfnamefont {M.}~\bibnamefont {Lobino}},\ }\href
{\doibase 10.1364/OE.23.001748} {\bibfield  {journal} {\bibinfo  {journal}
		{Optics Express}\ }\textbf {\bibinfo {volume} {23}},\ \bibinfo {pages} {1748}
	(\bibinfo {year} {2015})}\BibitemShut {NoStop}%
\bibitem [{\citenamefont {Symul}\ \emph {et~al.}(2011)\citenamefont {Symul},
	\citenamefont {Assad},\ and\ \citenamefont {Lam}}]{Symul:11}%
\BibitemOpen
\bibfield  {author} {\bibinfo {author} {\bibfnamefont {T.}~\bibnamefont
		{Symul}}, \bibinfo {author} {\bibfnamefont {S.~M.}\ \bibnamefont {Assad}}, \
	and\ \bibinfo {author} {\bibfnamefont {P.~K.}\ \bibnamefont {Lam}},\ }\href
{\doibase 10.1063/1.3597793} {\bibfield  {journal} {\bibinfo  {journal}
		{Applied Physics Letters}\ }\textbf {\bibinfo {volume} {98}},\ \bibinfo
	{pages} {23} (\bibinfo {year} {2011})}\BibitemShut {NoStop}%
\bibitem [{\citenamefont {S\"{o}nmez~Turan}\ \emph {et~al.}(2016)\citenamefont
	{S\"{o}nmez~Turan}, \citenamefont {Barker}, \citenamefont {Kelsey},
	\citenamefont {Boyle}, \citenamefont {Kerry},\ and\ \citenamefont
	{Baish}}]{Sonmez:16}%
\BibitemOpen
\bibfield  {author} {\bibinfo {author} {\bibfnamefont {M.}~\bibnamefont
		{S\"{o}nmez~Turan}}, \bibinfo {author} {\bibfnamefont {E.}~\bibnamefont
		{Barker}}, \bibinfo {author} {\bibfnamefont {J.}~\bibnamefont {Kelsey}},
	\bibinfo {author} {\bibfnamefont {M.}~\bibnamefont {Boyle}}, \bibinfo
	{author} {\bibfnamefont {M.}~\bibnamefont {Kerry}}, \ and\ \bibinfo {author}
	{\bibfnamefont {M.~L.}\ \bibnamefont {Baish}},\ }\href@noop {} {\emph
	{\bibinfo {title} {"Recommendation for the Entropy Sources Used for Random
			Bit Generation (Second DRAFT) NIST Special Publication 800-90B}}},\ \bibinfo
{organization} {National Institute of Standards and Technology} (\bibinfo
{year} {2016})\BibitemShut {NoStop}%
\bibitem [{\citenamefont {Dworkin}(2005)}]{Dworkin:05}%
\BibitemOpen
\bibfield  {author} {\bibinfo {author} {\bibfnamefont {M.}~\bibnamefont
		{Dworkin}},\ }\href@noop {} {\emph {\bibinfo {title} {Recommendation for
			Block Cipher Mode of Operation: The CMAC Mode for Authentication. NIST
			Special Publication 800-38B}}},\ \bibinfo {organization} {National Institute
	of Standards and Technology} (\bibinfo {year} {2005})\BibitemShut {NoStop}%
\bibitem [{NIS(2001)}]{NIST:01}%
\BibitemOpen
\href@noop {} {\emph {\bibinfo {title} {Specification for the Advanced
			Encryption Standard ( AES ) FIPS 197}}},\ \bibinfo {organization} {National
	Institute of Standards and Technology} (\bibinfo {year} {2001})\BibitemShut
{NoStop}%
\bibitem [{\citenamefont {Hsing}()}]{Hsing}%
\BibitemOpen
\bibfield  {author} {\bibinfo {author} {\bibfnamefont {H.}~\bibnamefont
		{Hsing}},\ }\href@noop {} {\emph {\bibinfo {title} {AES :: Overview ::
			OpenCores}}},\ \bibinfo {organization}
{https://opencores.org/project/tiny\_aes}\BibitemShut {NoStop}%
\bibitem [{\citenamefont {Bouda}\ \emph {et~al.}(2012)\citenamefont {Bouda},
	\citenamefont {Pivoluska},\ and\ \citenamefont {Plesch}}]{Bouda:12}%
\BibitemOpen
\bibfield  {author} {\bibinfo {author} {\bibfnamefont {J.}~\bibnamefont
		{Bouda}}, \bibinfo {author} {\bibfnamefont {M.}~\bibnamefont {Pivoluska}}, \
	and\ \bibinfo {author} {\bibfnamefont {M.}~\bibnamefont {Plesch}},\ }\href
{\doibase 10.1016/j.tcs.2012.07.030} {\bibfield  {journal} {\bibinfo
		{journal} {Theoretical Computer Science}\ }\textbf {\bibinfo {volume}
		{459}},\ \bibinfo {pages} {69} (\bibinfo {year} {2012})}\BibitemShut
{NoStop}%
\bibitem [{\citenamefont {Renner}(2008)}]{Renner:08}%
\BibitemOpen
\bibfield  {author} {\bibinfo {author} {\bibfnamefont {R.}~\bibnamefont
		{Renner}},\ }\href {\doibase 10.1142/S0219749908003256} {\bibfield  {journal}
	{\bibinfo  {journal} {International Journal of Quantum Information}\ }\textbf
	{\bibinfo {volume} {6}},\ \bibinfo {pages} {1} (\bibinfo {year}
	{2008})}\BibitemShut {NoStop}%
\bibitem [{\citenamefont {Rukhin}\ \emph {et~al.}(2010)\citenamefont {Rukhin},
	\citenamefont {Soto}, \citenamefont {Nechvatal}, \citenamefont {Miles},
	\citenamefont {Barker}, \citenamefont {Leigh}, \citenamefont {Levenson},
	\citenamefont {Vangel}, \citenamefont {Banks}, \citenamefont {Heckert},
	\citenamefont {Dray},\ and\ \citenamefont {Vo}}]{Rukhin}%
\BibitemOpen
\bibfield  {author} {\bibinfo {author} {\bibfnamefont {A.}~\bibnamefont
		{Rukhin}}, \bibinfo {author} {\bibfnamefont {J.}~\bibnamefont {Soto}},
	\bibinfo {author} {\bibfnamefont {J.}~\bibnamefont {Nechvatal}}, \bibinfo
	{author} {\bibfnamefont {S.}~\bibnamefont {Miles}}, \bibinfo {author}
	{\bibfnamefont {E.}~\bibnamefont {Barker}}, \bibinfo {author} {\bibfnamefont
		{S.}~\bibnamefont {Leigh}}, \bibinfo {author} {\bibfnamefont
		{M.}~\bibnamefont {Levenson}}, \bibinfo {author} {\bibfnamefont
		{M.}~\bibnamefont {Vangel}}, \bibinfo {author} {\bibfnamefont
		{D.}~\bibnamefont {Banks}}, \bibinfo {author} {\bibfnamefont
		{A.}~\bibnamefont {Heckert}}, \bibinfo {author} {\bibfnamefont
		{J.}~\bibnamefont {Dray}}, \ and\ \bibinfo {author} {\bibfnamefont
		{S.}~\bibnamefont {Vo}},\ }\href@noop {} {\emph {\bibinfo {title} {A
			statistical test suite for random and pseudorandom number generators for
			cryptographic applications. NIST Special Publication 800-22 Rev. 1a}}},\
\bibinfo {organization} {National Institute of Standards and Technology}
(\bibinfo {year} {2010})\BibitemShut {NoStop}%
\bibitem [{\citenamefont {McKay}\ and\ \citenamefont {Kelsey}()}]{github}%
\BibitemOpen
\bibfield  {author} {\bibinfo {author} {\bibfnamefont {K.}~\bibnamefont
		{McKay}}\ and\ \bibinfo {author} {\bibfnamefont {J.}~\bibnamefont {Kelsey}},\
}\href@noop {} {\emph {\bibinfo {title} {GitHub - usnistgov/SP800-90B
		\_EntropyAssessment}}},\ \bibinfo {organization}
{https://github.com/usnistgov/SP800-90B\_EntropyAssessment}\BibitemShut
{NoStop}%
\bibitem [{\citenamefont {Silverstone}\ \emph {et~al.}(2016)\citenamefont
	{Silverstone}, \citenamefont {Bonneau}, \citenamefont {O'Brien},\ and\
	\citenamefont {Thompson}}]{Silverstone:16}%
\BibitemOpen
\bibfield  {author} {\bibinfo {author} {\bibfnamefont {J.~W.}\ \bibnamefont
		{Silverstone}}, \bibinfo {author} {\bibfnamefont {D.}~\bibnamefont
		{Bonneau}}, \bibinfo {author} {\bibfnamefont {J.~L.}\ \bibnamefont
		{O'Brien}}, \ and\ \bibinfo {author} {\bibfnamefont {M.~G.}\ \bibnamefont
		{Thompson}},\ }\href {\doibase 10.1109/JSTQE.2016.2573218} {\bibfield
	{journal} {\bibinfo  {journal} {IEEE Journal of Selected Topics in Quantum
			Electronics}\ }\textbf {\bibinfo {volume} {22}},\ \bibinfo {pages} {390}
	(\bibinfo {year} {2016})}\BibitemShut {NoStop}%
\end{thebibliography}

%

\end{document}